\documentclass[review]{elsarticle}
\usepackage{hyperref}
\journal{Journal of \LaTeX\ Templates}

\usepackage{xcolor}
\begin{document}

\begin{frontmatter}

\title{First photometric investigation of V517 Cam combined with ground-based and TESS data}


\author[1]{Neslihan Alan}

\cortext[mycorrespondingauthor]{Corresponding author}
\ead{neslihan.alan@gmail.com}

\affiliation[1]{Fatih Sultan Mehmet Vakif University, Department of History of Science, 34664, Istanbul, Turkiye}

\author[2,3]{Fahri Ali{\c{c}}avu{\c{s}}}
\author[4]{Mehmet Alpsoy}

\affiliation[2]{Canakkale Onsekiz Mart University, Faculty of Science, Department of Physics, 17100, Canakkale, Turkiye}

\affiliation[3]{Canakkale Onsekiz Mart University, Astrophysics Research Center and Ulupınar Observatory, 17100, Canakkale, Turkiye}

\affiliation[4]{Akdeniz University, Institute of Graduate Studies in Science, Department of Space Sciences and Technologies, 07258, Antalya, Turkiye}

\begin{abstract}

The observations of eclipsing binary systems are of great importance in astrophysics, as they allow direct measurements of fundamental stellar parameters. By analysing high-quality space-based observations with ground-based photometric data, it becomes possible to detect these fundamental parameters with greater precision using multicolour photometry. Here, we report the first photometric analysis results of the V517 Cam eclipsing binary system by combining the Transiting Exoplanet Survey Satellite ({\it TESS}) light curve and new CCD observations in {\it BVRI} filters, obtained with a 60 cm robotic telescope (T60) at the T\"UB\.ITAK National Observatory. By means of photometric analyses, the masses and radii of the primary and secondary stars were carefully determined to be $M_{1}= 1.47\pm 0.06\,M_\odot$, $M_{2}= 0.79\pm0.05\,M_\odot$, and $R_{1}=1.43\pm 0.03\,R_\odot$, $R_{2}= 0.75\pm 0.04\,R_\odot$, respectively. Furthermore, the distance to V517 Cam was calculated to be $284\pm20$ pc. The overall age of the system is estimated to be around $63\pm15$ Myr. At this age, the primary component stands near the onset of its main-sequence evolution, near the ZAMS, whereas the secondary component remains in the pre-main-sequence evolutionary phase. To better understand the evolutionary status and nature of V517 Cam, the mass ratio and temperature values, obtained with relatively low sensitivity by photometric measurements, need to be confirmed by spectral analysis.

\end{abstract}

\begin{keyword}
\texttt{}Stars; binaries; eclipsing, Stars; fundamental parameters, Stars; evolution --- Stars: individual: V517 Cam ---  techniques: photometric
\end{keyword}
\end{frontmatter}

\section{Introduction}           
\label{sect:intro}

Eclipsing binary systems serve as crucial tools in astronomical research, offering significant insights into stellar structure, evolution, and galaxy dynamics. These systems allow for the direct determination of fundamental stellar parameters such as mass ($M$), radius ($R$), and luminosity ($L$) \citep{Eker24}. The accuracy of these measurements is greatly enhanced by high-quality photometric data from space telescopes like {\it TESS} \citep{Ricker15}. Improved accuracy in determining these parameters allows for the verification of theoretical models and the comparison of evolutionary predictions with observational data, leading to more accurate models. In particular, detached eclipsing binaries are suitable for examining the structure and evolution of single stars due to the minimal interaction between their components \citep{Eker24}. By using the fundamental parameters derived from observations of these binaries, the alignment between theoretical evolutionary models and observational data can be thoroughly evaluated. Additionally, detailed light curve analyses of detached eclipsing binaries that have not been previously studied provide important contributions to the literature by introducing new stellar parameters. In this context, the V517 Cam system, a detached binary for which only minima times have been documented and no light curve analysis has yet been conducted, is of particular interest.

V517 Cam was categorized by \citep{Kazarovets11} as an Algol-type eclipsing binary system. It has been reported that the system shows a single-line radial velocity variation using {\it Gaia} data \citep{Gaia22}. Additionally, \citet{Khalatyan24} used {\it Gaia} DR3 data (most notably the low-resolution XP spectra) to obtain the atmosphere parameters and masses of 217 million stars, including V517 Cam. The $\log g$, $\log T_{\rm eff}$, mass, and metallicity determined as mean for the V517 Cam system are 4.181$\pm$0.065 cgs, 3.782$\pm$0.007 K, 1.119$\pm$0.063 $M_\odot$, and -0.032$\pm$0.070 dex, respectively. Apart from these general surveys, which include stellar atmosphere model parameters and mass of the system, detailed spectral and photometric studies of the V517 Cam have not been performed until now. Therefore, the fundamental stellar parameters of the components of the system remain elusive. In order to reveal these parameters, the first light curve analysis of V517 Cam was performed using data from {\it TESS} and T60 observations. This analysis has yielded precise derivations of the fundamental stellar parameters of the components. The catalogue information of the V517 Cam system retrieved from AAVSO Variable Star Index (VSX)\footnote{https://www.aavso.org/vsx} and SIMBAD astronomical database\footnote{https://simbad.cds.unistra.fr/simbad} is listed in Table~\ref{tab:starlog}.

\begin{table}
\label{tab:starlog}
  \caption{General information for V517 Cam.}
  \begin{center} 
           \begin{tabular}{lcc}
\hline
Parameter & Value\\
\hline
${\rm RA}$ $(ep=J2000)^{a}$ & $12^{\rm h} 06^{\rm m} 51^{\rm s}.04$\\
${\rm DEC}$ $(ep=J2000)^{a}$ & $+77^{\circ} 18^{'} 57^{''}.00$\\
$V$ ${\rm (mag)}$ & $10.86\pm 0.06$ \\
$G$ ${\rm (mag)}$ & $10.705\pm 0.003$ \\
$J$ ${\rm (mag)}$ & $9.776\pm 0.023$ \\
$H$ ${\rm (mag)}$ & $9.488\pm 0.023$ \\
$K_{\rm S}$ ${\rm (mag)}$ & $9.407\pm 0.016$ \\
$\varpi$ (mas) & $3.7944\pm 0.0121$ \\
$P$ (day)$^{a}$ & 2.09223 \\
Type$^{a}$ & EA\\ 

 \hline
     \end{tabular}\\ 
$^{a}$ Represents data from VSX Search, all others were taken from SIMBAD. 
     \end{center}
\end{table}

The paper outlining the first light curve analysis findings for V517 Cam is structured as follows: Section 2 provides a description of the observational data and an overview of the methodology used to calculate updated light elements. In Section 3, the simultaneous analysis procedure for combining {\it TESS} data with ground-based photometric data is detailed. Section 4 then discusses the outcomes of this analysis and presents the fundamental parameters derived for the V517 Cam. Finally, Section 5 offers an in-depth discussion of the results and the evolutionary state of the system.

\section{Observational data}
\label{sect:Obs}

New multicolour CCD observations of V517 Cam were conducted over 222 nights from 12th February 2019 to 29th January 2022 using the 60 cm robotic telescope at the T\"UB\.ITAK National Observatory (TUG). The T60 telescope is controlled via the OCAAS open-source software, officially known as TALON (refer to \citet{Parmaksizoglu14}). Initially, the FLI ProLine 3041-UV CCD was used for observations until July 23, 2019, after which the Andor iKon-L 936 BEX2-DD camera was used. The FLI ProLine 3041-UV CCD has an image scale of 0.51 arcsec per pixel, giving a field of view (FOV) of 17.4 arcmin, whereas the Andor iKon-L 936 BEX2-DD camera provides an image scale of 0.456 arcsec per pixel, with a FOV of 15.6 arcmin.

V517 Cam observations were performed using Bessell {\it BVRI} filters (see \citet{Bessell90}). The exposure time of each $BVRI$ filter was set to 20, 10, 10, and 10 seconds, respectively. Calibration frames, including sky flats and bias frames, were taken periodically to correct for pixel-to-pixel variations in the CCD sensor. TYC 4550-1322-1 and TYC 4550-587-1 were used as comparison and check stars, respectively, to ensure accuracy and consistency in differential photometry. The $V$ filter image of the T60 telescope is shown in Figure 1. Along with ground-based observations, the light curve analysis includes data from {\it TESS}, which surveys most of the sky in sectors with 27.4 days of observation per sector.  {\it TESS} operates within a wavelength range of 600-1000 nm, providing broadband photometric data \citep{Ricker15}. The {\it TESS} data for the V517 Cam in Sector 14 were obtained using an exposure time of 120 seconds from July 18 to August 14, 2019. The system's {\it TESS} data are accessed via the Mikulski Archive for Space Telescopes (MAST)\footnote{https://archive.stsci.edu/} database. Pre-search Data Conditioning Simple Aperture Photometry light curves were used for the analysis, as described by \citet{Ricker15}. The mean error bar of the photometric data is around 0.1\%.

\begin{figure}[!ht]
\begin{center}
\includegraphics*[scale=0.46,angle=000]{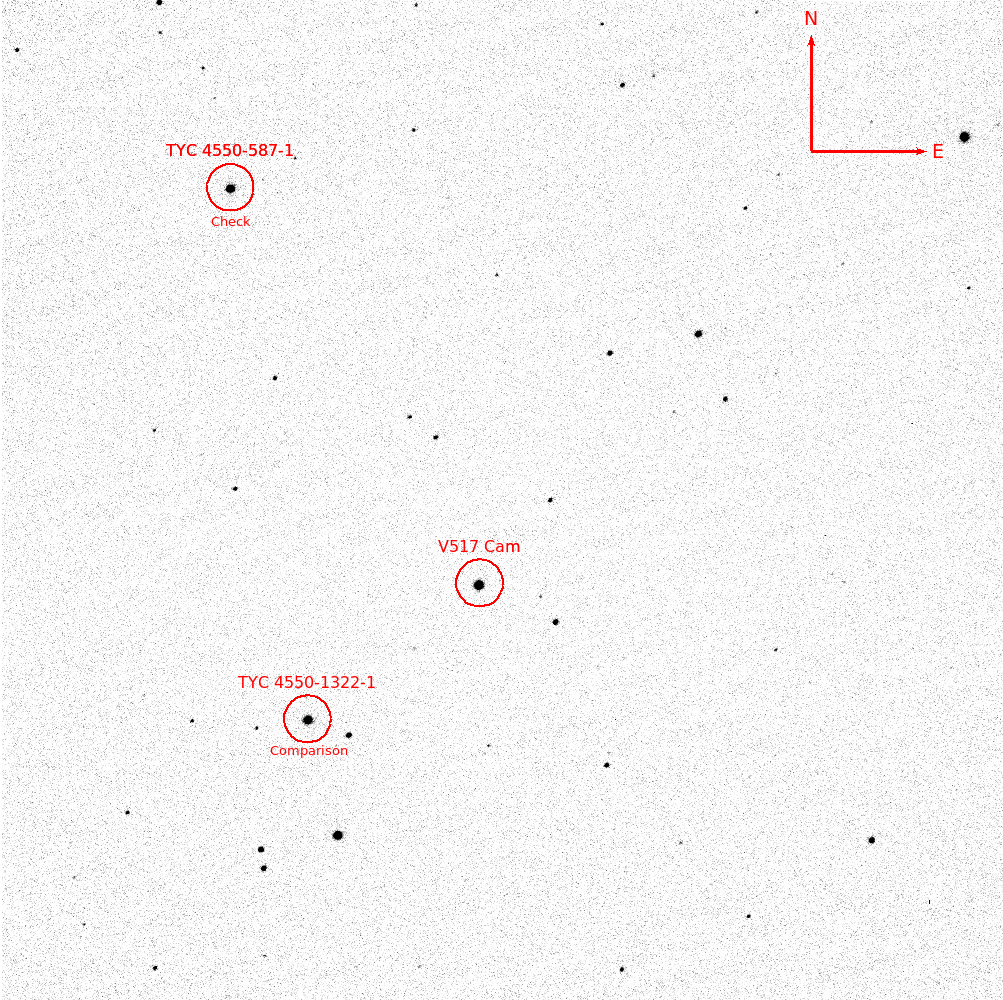}
\caption{$V$ filter image of the TUG-T60 telescope. Comparison, check stars, and V517 Cam are labelled on the image.} \label{CCD_image}
\end{center}
\end{figure}

The data reduction process for the T60 observations involved several main steps. The science frames were first subjected to bias and dark frame subtraction, then flat-field correction. The reduced CCD images obtained from these steps were later used to calculate the differential magnitudes for the target stars. This specific analysis was conducted using MYRaf software, the IRAF aperture photometry GUI tool created by \citet{Kilic16}. No notable light variations were observed in the comparison and check stars during the observation period. For the comparison and check star magnitude differences, the external uncertainties are found to be on mean 36 mmag in the filter $B$, 24 mmag in the filter $V$, 30 mmag in the filter $R$, and 30 mmag in the filter $I$. These uncertainties were determined based on the standard deviation of the differential magnitude changes between the comparison and check stars over the same night. The observational data were not transformed into the standard Bessell {\it BVRI} system, instead the differential magnitudes were used directly for light curve analysis.

For the V517 Cam eclipsing binary system, the minima times were calculated using {\it TESS} data. We obtained the primary minima times from the first, middle, and last parts of the {\it TESS} observations in sectors 14, 20, 21, 40, 41, 47, 48, 53, 60, 74, and 75. The secondary minima time was determined from the central section of the {\it TESS} dataset. This analysis yielded a total of 44 minima times from {\it TESS}. Additionally, minima times for V517 Cam from existing literature were gathered from the $Var Astro$ (O-C Gateway), yielding a total of 54 minima times. The minima times of the V517 Cam are listed in Table~\ref{tab:minimatimes}. We utilized these to investigate potential period variations in the system. Consequently, we derived updated light elements for V517 Cam by applying a linear fit to 54 minima times as shown in the equation below.

\begin{eqnarray}\label{eq1}
{\rm BJD(MinI)}=2458686.9660(1)+2^{\rm d}.0922270 (1)\times E
\end{eqnarray}
The values in parentheses represent the errors in the last numerical digit for the light elements in the equation.

\begin{table*}
\label{tab:minimatimes}
\setlength{\tabcolsep}{0.9pt}
\renewcommand{\arraystretch}{0.7}
\centering
  \caption{Minima times information of V517 Cam.}
  \vspace{0.5 cm}
   	\setlength{\tabcolsep}{5pt}
	\renewcommand{\arraystretch}{1}
    \begin{tabular}{cccc|cccc}
    \hline
    Minima Time & Error & Type & Reference & Minima Time & Error & Type & Reference \\     
    \hline
    \hline    
        2451581.7630 &  & p & \citep{Ote08} & 2459434.9389 & 0.0003 & s & This Study \\ 
        2455621.8537 & 0.0023 & p & $VarAstro^*$ & 2459435.9832 & 0.0002 & p & This Study \\ 
        2455981.7173 & 0.0030 & p & $VarAstro^*$ & 2459446.4440 & 0.0002 & p & This Study \\ 
        2456048.6666 & 0.0011 & p & $VarAstro^*$ & 2459582.4392 & 0.0002 & p & This Study \\ 
        2456688.8892 & 0.0002 & p & \citep{Nel15} & 2459596.0387 & 0.0004 & s & This Study \\ 
        2456814.4219 & 0.0012 & p & \citep{Hub15} & 2459597.0849 & 0.0002 & p & This Study \\ 
        2457131.3984 & 0.0031 & s & $VarAstro^*$ & 2459605.4539 & 0.0002 & p & This Study \\ 
        2457810.3229 & 0.0015 & p & $VarAstro^*$ & 2459611.7307 & 0.0002 & p & This Study \\ 
        2458155.5386 & 0.0060 & p & $VarAstro^*$ & 2459620.0995 & 0.0002 & p & This Study \\ 
        2458684.8731 & 0.0004 & p & This Study & 2459621.1456 & 0.0003 & s & This Study \\ 
        2458697.4269 & 0.0003 & p & This Study & 2459634.7448 & 0.0002 & p & This Study \\ 
        2458698.4727 & 0.0006 & s & This Study & 2459745.6325 & 0.0002 & p & This Study \\ 
        2458709.9805 & 0.0002 & p & This Study & 2459757.1397 & 0.0003 & s & This Study \\ 
        2458843.8831 & 0.0003 & p & This Study & 2459758.1862 & 0.0002 & p & This Study \\ 
        2458853.2980 & 0.0003 & s & This Study & 2459768.6475 & 0.0002 & p & This Study \\ 
        2458854.3445 & 0.0002 & p & This Study & 2459940.2099 & 0.0002 & p & This Study \\ 
        2458866.8978 & 0.0002 & p & This Study & 2459948.5788 & 0.0002 & p & This Study \\ 
        2458871.0821 & 0.0002 & p & This Study & 2459949.6249 & 0.0004 & s & This Study \\ 
        2458872.1291 & 0.0004 & s & This Study & 2459961.1323 & 0.0002 & p & This Study \\ 
        2458883.6353 & 0.0002 & p & This Study & 2460318.9023 & 0.0002 & p & This Study \\ 
        2458896.1887 & 0.0002 & p & This Study & 2460326.2270 & 0.0003 & s & This Study \\ 
        2459266.5119 & 0.0028 & p & $VarAstro^*$ & 2460327.2714 & 0.0002 & p & This Study \\ 
        2459392.0466 & 0.0002 & p & This Study & 2460337.7328 & 0.0002 & p & This Study \\ 
        2459402.5079 & 0.0002 & p & This Study & 2460346.1017 & 0.0002 & p & This Study \\ 
        2459403.5540 & 0.0004 & s & This Study & 2460354.4706 & 0.0002 & p & This Study \\ 
        2459417.1533 & 0.0002 & p & This Study & 2460355.5175 & 0.0003 & s & This Study \\ 
        2459421.3378 & 0.0002 & p & This Study & 2460367.0238 & 0.0002 & p & This Study \\ 
        \hline
    \end{tabular}
     \begin{description}
     \centering
 \item[ ] * https://var.astro.cz/en.
 \end{description}
\end{table*}

\section{Light Curve Analysis}
\label{sect:LightCurve}

A detailed analysis of the V517 Cam eclipsing binary star system’s light curve data was performed, utilizing a range of photometric filters, including normalized $BVRI$ and {\it TESS} data. The Wilson--Devinney (W--D) method \citep{WilsonDevinney71} was used in this analysis, and the uncertainties of the determined parameters were investigated by Monte-Carlo simulations \citep{Zola04, Zola10}. Since spectroscopic radial velocity measurements were not available for the V517 Cam system, we first determined $q$ using the comprehensive $q$-search method from light curve modelling to determine the most probable mass ratio. This technique, also known as the photometric mass ratio or $q$-search test, which aims to achieve a minimum $\chi^2$ value in iterations, is preferred in the absence of radial velocities of the binary star components \citep[see][]{Fa24, Fka22, Gurol05}. In the $q$-search, all parameters were left free and the $q$ value was fixed at each iteration with a step of 0.01 between 0 and 1. This investigation yielded low $\chi^2$ values in the range 0.3-0.6 and the initial value of $q$ was obtained around 0.5, considering the minimum $\chi^2$ for the system. The sum of the squared residuals corresponding to the $q$-values of the $q$-search is given in Figure 2. As seen from Figure 2, the $q$-search curve does not exhibit a well-defined sharp minimum. According to the $q$-search curve, the precision of the mass ratio is not very high. For more precise $q$ determination, high-resolution spectral data are essential.

\begin{figure}[!ht]
\begin{center}
\includegraphics*[scale=0.46,angle=000]{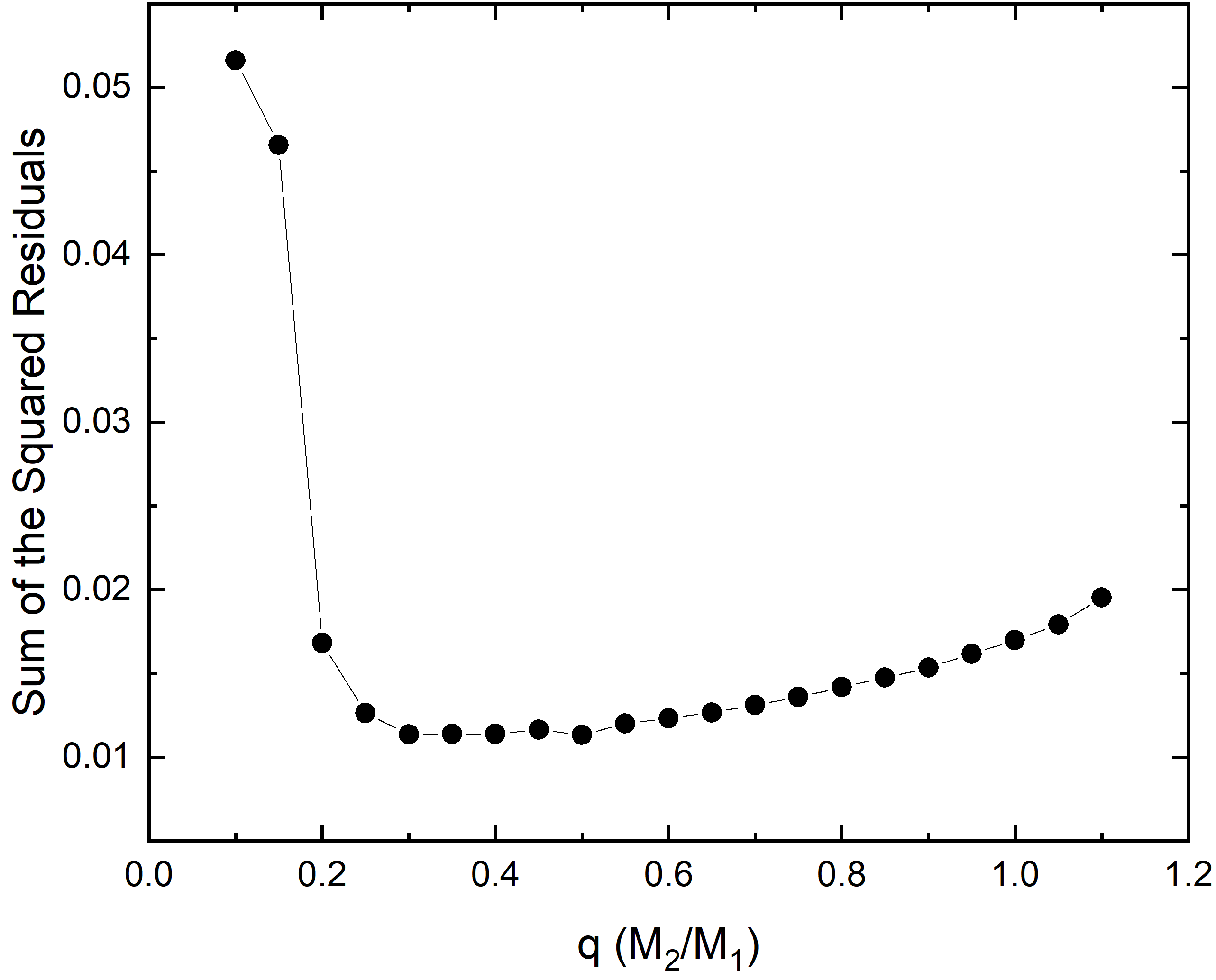}
\caption{Presentation of $q$-search results for V517 Cam.} \label{q_search}
\end{center}
\end{figure}

For the analysis with the W--D code, the approach used includes a mixture of fixed parameters grounded in theoretical models and previous research, along with parameters that are iteratively adjusted by consecutive iterations. The fixed and adjusted parameters are presented as follows. The unreddened colour index of V517 Cam was calculated as $(B-V)_{\rm 0}=0.413\pm0.02$ mag, using a colour excess $E_{\rm d}(B-V)=0.055$ mag derived from \citet{Schlafly11}’s Galactic dust map and a colour index $B-V = 0.468\pm0.02$ mag taken from the Tycho-2 catalogue data \citep{Hog00}. The primary component’s temperature was set to $T_{\rm 1,eff}=6588$ K, in consistency with the $B-V$ colour index obtained by \citet{Eker20} using the astrophysical parameters of the main-sequence stars (Solution-I). Furthermore, a temperature of 6053 K for the primary component was determined by \citet{Khalatyan24} using {\it Gaia} DR3 data. Fixing this temperature value for the primary component, a second light curve solution of the system was performed (Solution-II). Due to the nature of the observed light variations in both solutions, the simultaneous analysis of the ground-based $BVRI$ light curves and {\it TESS} data was done using mode 2 for detached binary systems. We adopted the root-mean-square limb darkening law and retrieved the coefficients of limb darkening from the \citet{Vanhamme93} tables based on the filter wavelengths and temperatures of the V517 Cam components. In a convective atmosphere ($T_{\rm eff}<7200$ K) approach, the bolometric gravity-darkening exponents of the components were set to 0.32 from \citet{Lucy67} and the bolometric albedo was held constant at 0.5 obeying \citet{Rucinski69}. We assume that both components are in rotation synchronously ($F_{\rm 1}=F_{\rm 2}=1$). The orbital eccentricity parameter $e$ was initially left free in the model and its value approached zero after iterations. Thus, it was fixed at $e=0$ in the subsequent analyses. The fact that the secondary minima of V517 Cam is in phase 0.5, that there is no obvious asymmetry in the observed light curve, and that the ascent and descent durations are the same for the primary and secondary minima all support a circular ($e=0$) orbit. The orbital inclination ($i$), the surface temperature of secondary component ($T_{\rm 2,eff}$), the mass ratio ($q$), the dimensionless surface potential of primary and secondary components ($\Omega_{1,2}$), phase shift, and fractional luminosity of primary component ($L_{1}$), were considered as adjustable parameters in the analysis. The possibility of a third body contributing to the total light was evaluated by taking into account as a free parameter $l_{3}$ in the analysis. No third light effect was detected for the V517 Cam. 
The resulting model parameters for Solutions I and II are given in detail in Table~\ref{tab:lcparameters}. Additionally, for Solution-I, the comparison between the observed and calculated light curves is shown in Figure 3, and the Roche geometry of the system is plotted in Figure 4.

\begin{table*}
\label{tab:lcparameters}
\begin{center}
\centering
\caption{Results of the light curve analyses of V517 Cam. Subscripts 1, 2, and 3 refer to the primary, secondary, and third components, respectively. $^a$ denotes fixed parameters.}
\begin{tabular}{lrrr}
\hline
 Parameter			 			               & Solution-I	 & Solution-II	\\	
\hline
$T_{0}$ {\rm (BJD+2400000)}                     & \multicolumn{2}{c}{58686.9660} \\
$P_{\rm orb}$ (days)                            & \multicolumn{2}{c}{2.092227}  \\
$i$ ($^{\rm o}$)	       	                    & 89.465 $\pm$ 0.014 & 89.418 $\pm$ 0.014  \\	
$T$$_{\rm 1,eff}$$^a$ (K)                 		        & 6588 $\pm$ 120 & 6053 $\pm$ 100	\\	
$T$$_{\rm 2,eff}$ (K)    	          		            & 4354 $\pm$ 150 & 4155 $\pm$ 125		\\
$e$	         	               		            & 0.000	& 0.000	  \\
$\Omega$$_{1}$		           	                & 6.837 $\pm$ 0.024 & 6.806 $\pm$ 0.023   	  \\
$\Omega$$_{2}$		            	            & 7.700 $\pm$ 0.071 & 7.651 $\pm$ 0.071	  \\
Phase shift             	  	                & 0.0000 $\pm$ 0.0001 & 0.0000 $\pm$ 0.0001	  \\
$q$                     	  	                & 0.535 $\pm$ 0.030 & 0.531 $\pm$ 0.029  	  \\
$r$$_{\rm 1}^*$ (mean)                          & 0.1587 $\pm$ 0.0009 & 0.1597 $\pm$ 0.0009 \\
$r$$_{\rm 2}^*$ (mean)                          & 0.0828 $\pm$ 0.0006 & 0.0828 $\pm$ 0.0006 \\
$L$$_{\rm 1}$/($L$$_{1}$+$L$$_{2}$) ($TESS$) 	& 0.942 $\pm$ 0.001 & 0.941 $\pm$ 0.001   \\
$L$$_{\rm 1}$/($L$$_{1}$+$L$$_{2}$) ($B$)   	& 0.975 $\pm$ 0.006 & 0.973 $\pm$ 0.006   \\
$L$$_{\rm 1}$/($L$$_{1}$+$L$$_{2}$) ($V$)   	& 0.967 $\pm$ 0.006 & 0.965 $\pm$ 0.006   \\
$L$$_{\rm 1}$/($L$$_{1}$+$L$$_{2}$) ($R$)  	    & 0.955 $\pm$ 0.005 & 0.953 $\pm$ 0.005   \\
$L$$_{\rm 1}$/($L$$_{1}$+$L$$_{2}$) ($I$)   	& 0.942 $\pm$ 0.005 & 0.942 $\pm$ 0.005   \\
$L$$_{\rm 2}$/($L$$_{1}$+$L$$_{2}$) ($TESS$)    & 0.058 $\pm$ 0.001 & 0.059 $\pm$ 0.001 	 \\
$L$$_{\rm 2}$/($L$$_{1}$+$L$$_{2}$) ($B$)   	& 0.025 $\pm$ 0.002 & 0.027 $\pm$ 0.002	 \\
$L$$_{\rm 2}$/($L$$_{1}$+$L$$_{2}$) ($V$)  	    & 0.033 $\pm$ 0.002 & 0.035 $\pm$ 0.002 	 \\
$L$$_{\rm 2}$/($L$$_{1}$+$L$$_{2}$) ($R$)   	& 0.045 $\pm$ 0.002 & 0.046 $\pm$ 0.002	 \\
$L$$_{\rm 2}$/($L$$_{1}$+$L$$_{2}$) ($I$)   	& 0.058 $\pm$ 0.002 & 0.058 $\pm$ 0.002	 \\
$l$$_{\rm 3}$ ($TESS$)                        	& 0.0 & 0.0	        	  \\
 \hline
\end{tabular}
     \end{center}
     \begin{description}
     \centering
 \item[ ] * fractional radii calculated from the geometric mean $r_{\rm mean}=(r_{\rm pole} \times r_{\rm side} \times r_{\rm back})^{1/3}$.
 \end{description}
\end{table*}

\begin{figure}[!ht]
\begin{center}
\includegraphics*[scale=0.5,angle=000]{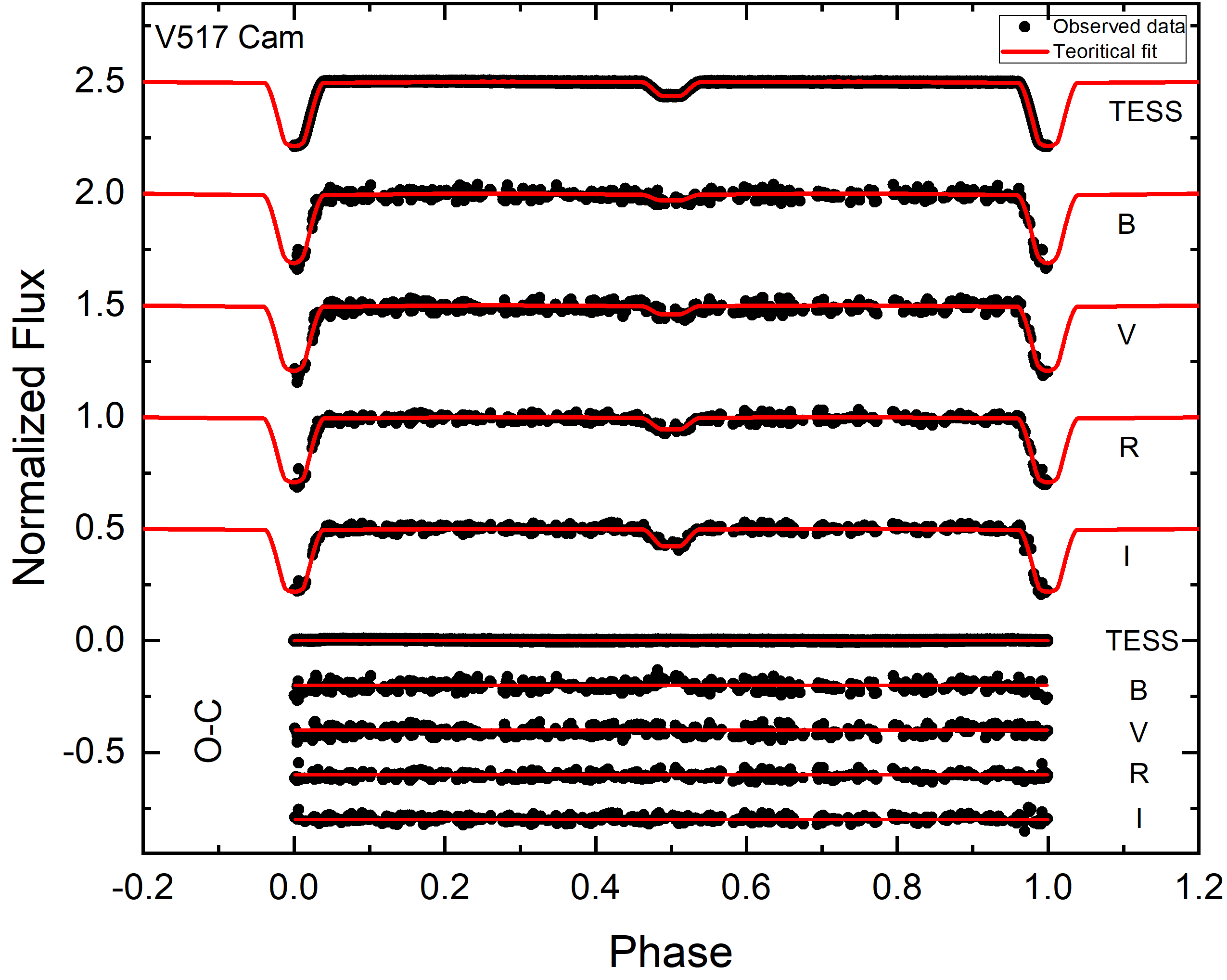}
\caption{A comparison of the theoretical light curves (red line) with the observational data (black dot) of the V517 Cam. The bottom panel represents the residuals.} \label{LC_fit}
\end{center}
\end{figure}

\begin{figure}[!ht]
\begin{center}
\includegraphics*[scale=.50,angle=000]{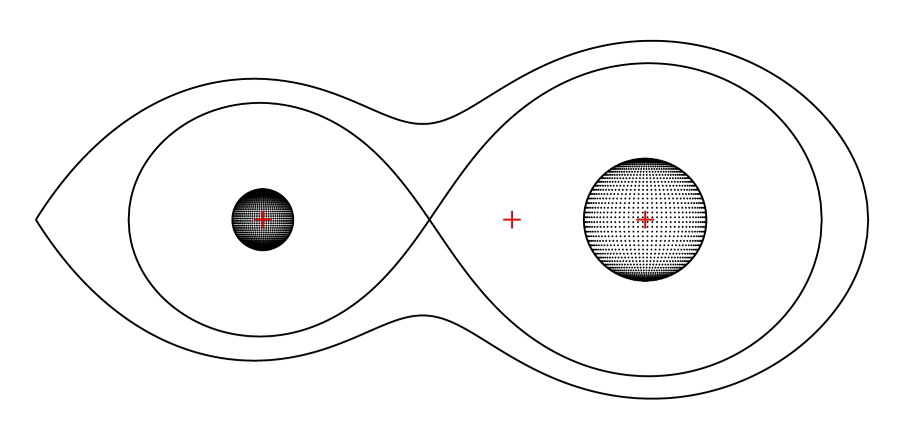}
\caption{Roche geometry generated using the parameters of the best light curve model of V517 Cam.} \label{fig:Roche_geometry}
\end{center}
\end{figure}

\section{Estimated Fundamental Parameters and Evolutionary Status}
\label{sect: Absolute Parameters}

Based on the findings from both Solution I and II photometric analyses,  two different approaches were used to determine the fundamental parameters for the components of the V517 Cam. In the first approach, supposing that the primary component is a main-sequence star with an effective temperature $T_{\rm 1,eff}=6588$ K, the mass of the primary component was determined as $M_{1}=1.47\,M_{\odot}$ from the correlation between mass and $T_{\rm eff}$ for main-sequence stars given by \citet{Eker20} for V517 Cam. The secondary component mass was calculated from the mass ratio yielded from the light curve analysis. The radii of the components were ascertained using the mean fractional radii in Table~\ref{tab:lcparameters} and the semi-major axis calculated by Kepler's third law. The resulting temperature and radius values were used in the common formula $L=4\pi R^{2} \sigma T_{\rm eff}^{4}$ to calculate the luminosities of the components. Then, bolometric magnitudes are estimated using these luminosities. The luminosities and bolometric magnitudes of the components were estimated by adopting standard solar values ($T_{{\rm eff, \odot}} = 5777$ K, $M_{{\rm bol},\odot} = 4.74$ mag) and using bolometric corrections from \citet{Eker20}. The surface gravity values were calculated as follows for the primary and secondary components, respectively: log\,$g_{1} = 4.29 \pm 0.02$ and log\,$g_{2}=4.59 \pm 0.04$ in cgs units for V517 Cam. The second approach is based on the values of temperature (6053 K) and mass (1.12 $M_\odot$) estimated by \citet{Khalatyan24} from {\it Gaia} DR3 data, and then the fundamental parameters of the components are recalculated by following the steps outlined in the previous lines. The estimated fundamental parameters from both approaches are listed in detail in Table~\ref{tab:estparameters}.

The dust map of \citet{Schlafly11} was used to obtain the $V$-band extinction in the direction of V517 Cam. The Galactic coordinates ($l = 126^{\circ}.09$, $b =+39^{\circ}.54$) of V517 Cam were determined as $A_{\infty}(b)=0.232\pm0.004$ mag in the direction of the system via the dust extinction calculation tool on the NASA/IPAC website\footnote{https://irsa.ipac.caltech.edu/applications/DUST/}. 
The reduced extinction ($A_{\rm d}(b)$) between the Sun and V517 Cam was calculated using the \citet{Bahcall1980} relation:

\begin{equation}
A_{d}(b)=A_{\infty}(b)\Biggl[1-\exp\Biggl(\frac{-\mid
d\times \sin(b)\mid}{H}\Biggr)\Biggr]
\label{eq:indirgeme}
\end{equation}
here $d$ and $b$ are the distance of the V517 Cam and the Galactic latitude, respectively. $H$ is the scale height for the interstellar
dust which is adopted as 125 pc \citep{Marshall2006} and $A_{\infty}(b)$ and $A_{\rm d}(b)$ are the total absorption for the dust map and the distance to the V517 Cam, respectively \citep[see also,][]{Bilir2008a, Bilir2008b, Eker09}. In addition, the colour excess ($E_{\rm d}(B-V)$) in the direction of V517 Cam was computed using the relation $E_{\rm d}(B-V)=A_{\rm d}(b)~/~3.1$. By substituting the distance of the V517 Cam ($264\pm 1$ pc) estimated from the trigonometric parallax ($\varpi=3.7944\pm0.0121$ mas) from \citet{Gaia23} catalogue into Equation \ref{eq:indirgeme}, the colour excess and $V$-band extinction were obtained as $E_{\rm d}(B-V)=0.055\pm0.001$ and $A_{\rm d}(V)=0.0172\pm 0.002$ mag, respectively.

The distance of V517 Cam was obtained as $284\pm20$ pc based on the interstellar extinction given in Table~\ref{tab:estparameters}, the apparent magnitude of the system, the component light ratios listed in Table~\ref{tab:lcparameters}, and the values $BC_{1}$= $0.074$ mag and $BC_{2}$= $-0.655$ mag calculated according to \citet{Eker20}. The resulting fundamental stellar parameters offer a detailed insight into the evolutionary state and age of binary component stars. The MESA Isochrones \& Stellar Tracks (MIST) framework \citep{{Choi16},{Dotter16},{Paxton11},{Paxton13},{Paxton15},{Paxton18}} was used to investigate the evolutionary status of the components. The evolutionary tracks that best represent the calculated fundamental parameters in the Hertzsprung Russell diagram were obtained with a metallicity of $Z$\,=\,0.027 $\pm$ 0.002 (shown in Figure 5). Examining the radius age ($\log R-age$) diagrams of the components, the best age agreement for the massive component was found to be 63\,$\pm$\,15 Myr (depicted in Figure 6 top panel). This age indicates that the massive component is at the very early stages of main-sequence evolution, near the Zero Age Main Sequence (ZAMS), while the low-mass component has not yet settled onto the main sequence and is still in the pre-main-sequence evolutionary phase. According to the parameters obtained from the Solution-II, the age of the system is 4.70$\pm$0.30 Gyr and the best metallicity fit is $Z$\,=\,0.017 $\pm$ 0.002. The concordance of the evolution models with the calculated parameters is given in Figure 5 along with the results of Solution-I, and in the bottom panel of Figure 6.

\begin{table*}
\label{tab:estparameters}
\begin{center}
\centering
\caption{The estimated fundamental stellar parameters for V517 Cam.}
\begin{tabular}{lrr}
\hline
  Parameter 			              & Solution-I & Solution-II			\\	
\hline
$M$$_{1}$ ($M_\odot$)	          	&1.47 $\pm$ 0.06 &1.12 $\pm$ 0.06   \\	
$M$$_{2}$ ($M_\odot$)	         	&0.79 $\pm$ 0.05 &0.59 $\pm$ 0.04   \\
$R$$_{1}$ ($R_\odot$)	          	&1.43 $\pm$ 0.03 &1.31 $\pm$ 0.03   \\
$R$$_{2}$ ($R_\odot$)		  		&0.75 $\pm$ 0.04 &0.68 $\pm$ 0.03   \\
$a$ ($R$$_{\odot}$)        	   	    &9.02 $\pm$ 0.12 &8.23 $\pm$ 0.11   \\
$\log L$$_{1}$ ($L_\odot$)		  	&0.54 $\pm$ 0.03 &0.32 $\pm$ 0.03  \\
$\log L$$_{2}$ ($L_\odot$)		  	&-0.74 $\pm$ 0.06 &-0.90 $\pm$ 0.06    \\
$\log g$\,$_{1}$ (cgs)              &4.29 $\pm$ 0.02 &4.25 $\pm$ 0.02   \\
$\log g$\,$_{2}$ (cgs)              &4.59 $\pm$ 0.04 &4.55 $\pm$ 0.04   \\
$M_{\rm Bol, 1}$ (mag)              &3.39 $\pm$ 0.13 &3.94 $\pm$ 0.14   \\
$M_{\rm Bol, 2}$ (mag)	 	        &6.60 $\pm$ 0.27 &7.00 $\pm$ 0.28  \\
$M_{\rm V, 1}$ (mag)	         	&3.46 $\pm$ 0.15 &3.97 $\pm$ 0.16   \\
$M_{\rm V, 2}$ (mag)	          	&7.25 $\pm$ 0.29 &7.86 $\pm$ 0.30  \\
$A_{\rm V,d}$ (mag)	         	    &0.177 $\pm$ 0.020 &0.158 $\pm$ 0.020 \\
$d$ (pc)                            &284 $\pm$ 20  &226 $\pm$ 18    \\
 \hline
\end{tabular}
     \end{center}
\end{table*}

\begin{figure}[!ht]
\begin{center}
\includegraphics*[scale=.5,angle=000]{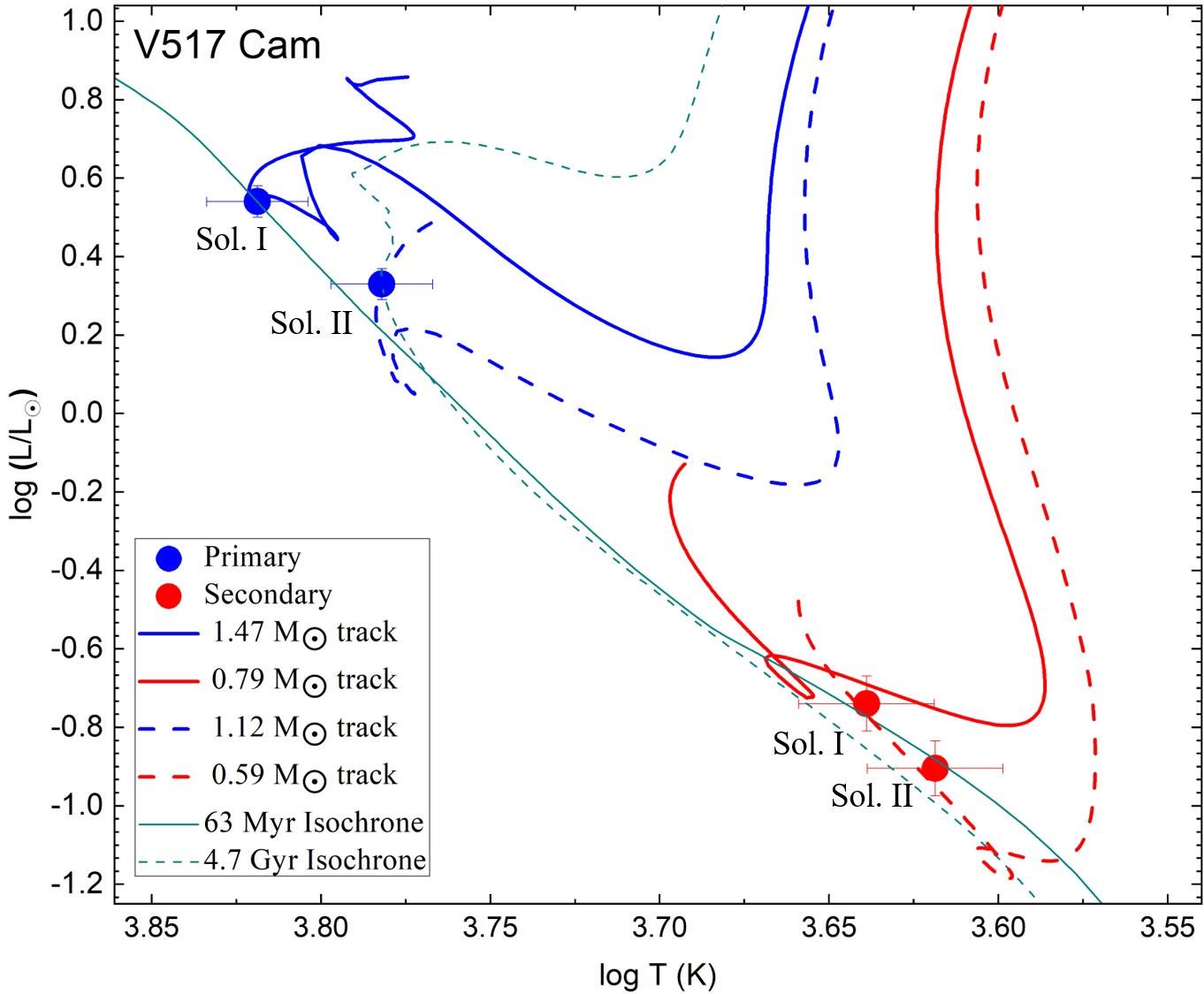}
\caption{The positions of the primary (blue dot) and secondary (red dot) components of V517 Cam in the log\,$L$ - log\,$T$ plane. Evolutionary tracks with respect to the metallicities $Z = 0.027$ and $Z = 0.017$ are indicated by the blue and red lines for the primary and secondary components, respectively. Theoretical evolution curves were calculated with the MESA code.} \label{LogL_Te}
\end{center}
\end{figure}

\begin{figure}[!ht]
\begin{center}
\includegraphics*[scale=.41,angle=000]{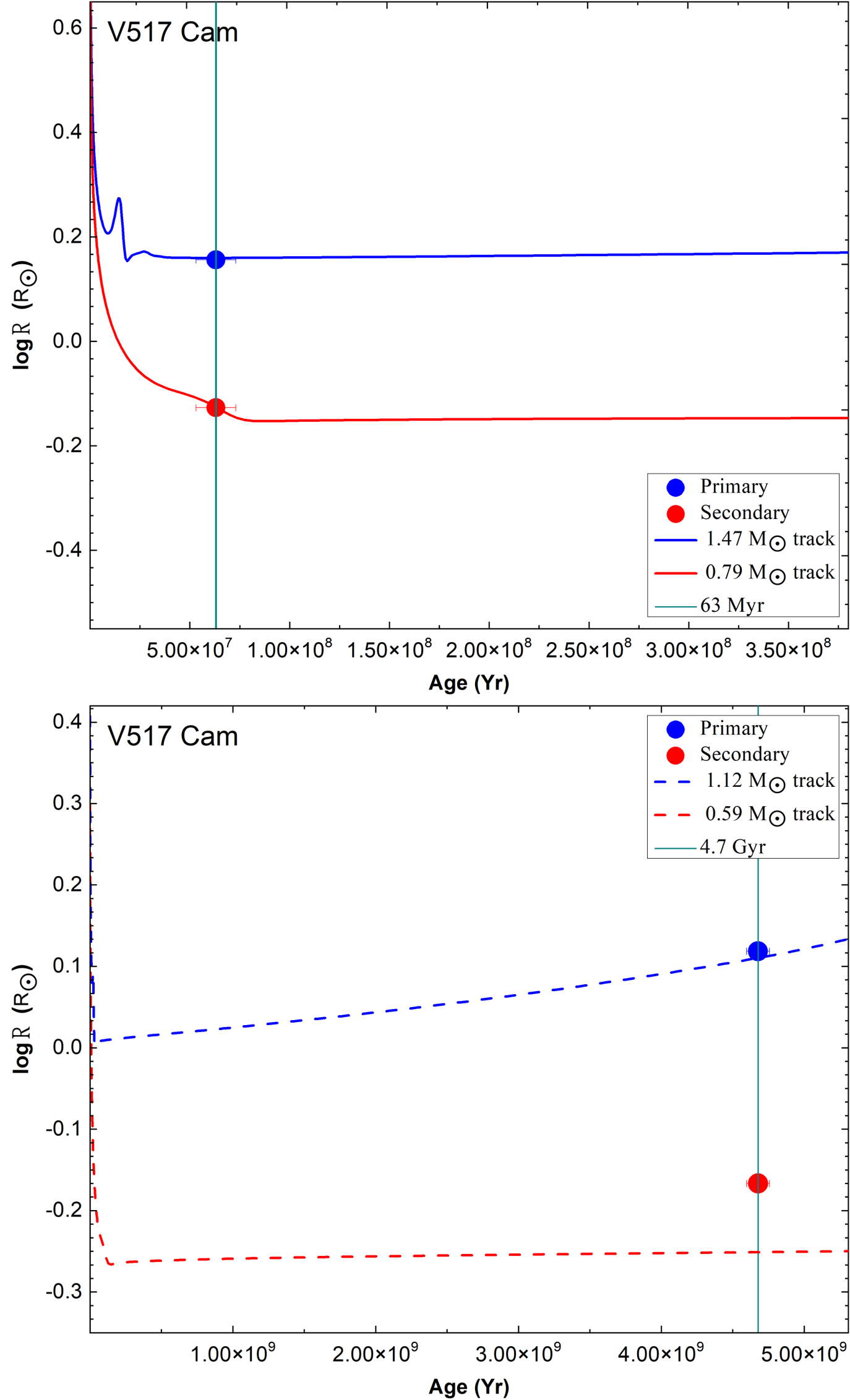}
\caption{The positions of the primary (blue dot) and secondary (red dot) components of V517 Cam in the log\,$R$ - Age plane. Evolutionary tracks for metallicities of $Z=0.027$ and $Z = 0.017$ are indicated by the blue and red lines for the primary and secondary components, respectively.} \label{R_age}
\end{center}
\end{figure}

\section{Discussion and Conclusions}
\label{sect:discussion}

We presented the first light curve analysis results of high-quality {\it TESS} data and CCD multicolour observational data sets in {\it BVRI} filters to obtain the fundamental parameters of the V517 Cam. Moreover, by using the minima measured from {\it TESS} data and collected from the literature, new light elements of the system were computed. In this study, two solutions were performed in which the temperatures of the primary component obtained from two different methods were kept fixed. The Solution-I was based on the widely used $B-V$ temperature determination method in the literature. In Solution-II, the temperature value was used, derived by \citet{Khalatyan24} from low-resolution $Gaia$ spectral data. The mass, radii, temperature, surface gravity, and luminosity parameters obtained from Solution-I are relatively larger compared to Solution-II based on the values given by \citet{Khalatyan24}. The main reason for this may be that the binarity situation was not considered and the data precision was not sufficient in the study of \citet{Khalatyan24}. In some cases, {\it Gaia} temperatures differ significantly from measurements using high-resolution spectra \citep[see][]{Kahraman23, Fa24, Yucel25}. Although there are similar uncertainties in the temperature values obtained from the $B-V$ colour index, the distance values from Solution-I are more consistent with the $Gaia$ distance. In addition, the findings obtained from Solution-I are more interesting due to the position of the components in the H-R diagram and the age of the system. For this reason, Solution-I is preferred as the main analysis in this study. It needs to be noted that the $q$ values obtained from both solutions with two different temperatures are almost the same. This indicates that the light curve analysis is not highly sensitive to the assumed primary star temperature (e.g. \citet{Zhang13}). Also, based on the $q$-search curve, the precision of the mass ratio is not very high (see Figure 2). Consequently, there are uncertainties in the temperatures and $q$-values obtained from photometric analysis. To accurately determine these parameters, analyses with high-resolution spectral observation data are needed.  

As a result of the solution-I, the masses of the components were found to be $M_{1}$\,=\,1.47\,$\pm$\,0.06\,$M_\odot$ and $M_{2}$\,=\,0.79\,$\pm$\,0.05\,$M_\odot$. The radii of the primary and secondary components of the system are determined respectively as follows: $R_{1}$\,=\,1.43\,$\pm$\,0.03\,$R_\odot$ and $R_{2}$\,=\,0.75\,$\pm$\,0.04\,$R_\odot$. Additionally, the photometric analysis shows that the distance of V517 Cam is about 284 $\pm$ 20 pc, which is consistent with the {\it Gaia}-DR3 distance of 264 $\pm$ 1 pc \citep{Gaia23} within uncertainties. The temperatures of the components of the system were estimated by light curve analysis of the V517 Cam as $T_{\rm 1,eff}\,=\,6588$ K and $T_{\rm 2,eff}\,=\,4354$ K. We compared the obtained temperatures with the temperatures in the table given by \citep{Eker20} for main sequence stars and found that the spectral types for the primary and secondary components are F4 and K5, respectively.

The distance of $284\pm 20$ pc calculated using photometric methods for V517 Cam in this study is slightly larger than the trigonometric parallax distance given in the {\it Gaia} DR3 catalogue ($d=264\pm 1$ pc), which was determined using the relation $d(pc)=1/\varpi$. Although the distances calculated using the photometric method and the trigonometric parallax method are consistent within uncertainties, the distance of the system was also tested using an alternative photometric method. The $J$-band apparent and absolute magnitudes of V517 Cam were calculated as $J_{\rm 0} = 9.727$ and $M_{\rm J} = 2.459$, mag, respectively, using the photometric calibration provided in Eq. 6 of \citet{Bilir2008a}, which is sensitive to infrared photometric data. Using the distance modulus, the system's distance depending on the infrared magnitudes was determined as $284\pm5$ pc. It is noteworthy that the two distances calculated based on optical and infrared magnitudes have exactly the same value, but both are larger than the {\it Gaia} distance. This may be an effect of the binary nature of the system in terms of {\it Gaia} satellite data. Moreover, the distance calculated from Solution-II with the parameters obtained from {\it Gaia} data is incompatible with the {\it Gaia} distance, it is smaller.

\citet{Griffin85} reported that double-lined spectroscopic binaries on the main sequence show a net peak around $q$ $\approx$ 1. \citet{Eker14} also provided a median mass ratio value $q=0.931$ for detached binary systems with the primary component of spectral type F. The mass ratio $q=0.535\pm0.030$ of V517 Cam obtained in this work differs from the mean values given by \citet{Griffin85} and \citet{Eker14}. The number of detached systems with large mass differences, such as V517 Cam, is relatively small. In \citet{Eker18}'s study of 293 double-lined detached binary systems, there are only 30 systems with a mass ratio of less than 0.6. This makes systems as limited in number such as V517 Cam important for revealing the evolutionary processes of components with quite different masses. The estimated mass values for each component were fitted to the mass-luminosity correlation in the relevant mass range provided by \citet{Eker18} for the main-sequence stars. The mean luminosity values for the primary and secondary components were calculated according to this correlation as $\log L_1=0.73\,L_{\odot}$ and $\log L_2=-0.61\,L_{\odot}$, respectively. The mean luminosity values calculated from \citet{Eker18} equations were found to be larger than the observational luminosity values (for Solution-I; $\Delta L_1$ = 0.19$\,L_{\odot}$, $\Delta L_2$ = 0.13$\,L_{\odot}$). These values demonstrate that the components of V517 Cam exhibit lower luminosity values compared to the stars within the $1.05<M/M_\odot \leq 2.40$ (intermediate-mass) and $0.72<M/M_\odot \leq 1.05$ (low mass) mass range reported by \citet{Eker18}. Furthermore, it suggests that the system's age is younger than the mean age of main-sequence stars. Similar cases have also been highlighted in other studies on detached binary stars, such as those by \citet{Alan23}, \citet{Alan24}, \citet{Alicavus22}, and \citet{Yucel24}.

Investigating the evolution of the system's components, the best age match for the components within uncertainties for Solution-I was around 63 Myr. This finding, as previously mentioned, indicates that the components are relatively young and located near the ZAMS. Specifically, the secondary component is in the pre-main-sequence phase and is expected to settle onto the ZAMS in approximately the next 20 Myr. It is crucial to examine this scenario in detail through future spectroscopic studies. For such late spectral types of stars, there are significant differences in radii and temperatures compared to the values expected from the model due to magnetic activity, as often reported in previous studies \citep{{Tor08}, {Sta09}, {Fa19}}. Therefore, age determination for binary stars with such components using the properties of the primary component yields more accurate results. Investigations of systems containing such components are important to better understand the rotational effects on magnetic activity and to understand the impact of the mass-radius discrepancy in the evolutionary process.

The components of the V517 Cam were examined in this work using very precise photometric data. This enabled us to establish the fundamental stellar parameters for the components. Detached binary systems for example V517 Cam offer a rare opportunity to directly quantify stellar masses, radii, and luminosities, which are typically difficult to measure precisely for single stars. Altogether, the V517 Cam eclipsing binary system represents a valuable laboratory for expanding our knowledge of stellar evolution and interactions of binary stars. It is essential to obtain the radial velocity curves of both components to increase the precision of computing the mass ratio and the fundamental stellar parameters. This requires that spectroscopic observations be performed on the V517 Cam. Following future observations, analyses combining spectroscopic data with photometric measurements could provide a better understanding of the nature of the system with more precise findings.

\section*{Acknowledgements}
We would like to thank the anonymous referee for insightful and constructive suggestions and comments that significantly improved the manuscript. This research was supported by the Scientific Research Projects Coordination Unit of Istanbul University. Project number: 37903. We would like to thank T\"UB\.ITAK National Observatory (TUG) for providing partial support for the use of the T60 telescope through project number 18BT60-1324. We also thank the observers and technical staff at TUG for their assistance before and during the observations. Our special thanks to Sel\c{c}uk Bilir for his useful contributions.

This study utilized NASA’s (National Aeronautics and Space Administration) Astrophysics Data System and SIMBAD Astronomical Database operated by CDS, Strasbourg, France, and the NASA/IPAC Infrared Science Archive operated by the Jet Propulsion Laboratory, California Institute of Technology under contract with the National Aeronautics and Space Administration. This research also utilized data from the European Space Agency (ESA) mission $Gaia$\footnote{https://www.cosmos.esa.int/gaia}, processed by the $Gaia$ Data Processing and Analysis Consortium (DPAC)\footnote{https://www.cosmos.esa.int/web/gaia/dpac/consortium}. Funding for DPAC was provided by national institutions, mainly those participating in the Gaia Multilateral Agreement. The {\it TESS} data presented in this paper were obtained from the  Mikulski Archive for Space Telescopes (MAST). The NASA Explorer Program funds the {\it TESS} mission.


\end{document}